\documentstyle[10pt,epsf,epsfig,dp_delphititle,oldlfont,pennames,hhline,
multicol,array,amstex,color]{dp_delphi}
\makeindex
\def\DpPaperGroup{EP}
\def\DpPaperRef{2000-095}
\def\DpDate{17 July 2000}
\def\DpAuthors{DELPHI Collaboration}
\def\DpSubmit{(Phys. Lett. B496(2000)59)}
\def\DpTitle{{Search for supersymmetric partners of top and 
        bottom quarks at \boldmath $\sqrt{s} = 189$ GeV}}
\def\DpComment{ }
\def\DpEMail{ }
\def\leqsim{\mathbin{\;\raise1pt\hbox{$<$}\kern-8pt\lower3pt\hbox{$\sim$}\;}}
\def\geqsim{\mathbin{\;\raise1pt\hbox{$>$}\kern-8pt\lower3pt\hbox{$\sim$}\;}}

\def\MXN#1{\mbox{$ M_{\tilde{\chi}^0_#1} $}}

\def\XN#1{\mbox{$ \tilde{\chi}^0_#1 $}}

\newcommand{\stq} {\mbox{$ \tilde {\mathrm t} $}}

\newcommand{\sbq} {\mbox{$ \tilde {\mathrm b} $}}

\newcommand{\Zn} {\mbox{$ {\mathrm Z}^0 $}}

\newcommand{\Wp} {\mbox{$ {\mathrm W}^+ $}}
\newcommand{\Wm} {\mbox{$ {\mathrm W}^- $}}

\newcommand{\ee} {\mbox{$ {\mathrm e}^+ {\mathrm e}^- $}}
\newcommand{\eeto} {\mbox{$ {\mathrm e}^+ {\mathrm e}^- \to $}}

\newcommand{\MeV} {\mbox{$ {\mathrm{MeV}} $}}
\newcommand{\MeVc} {\mbox{$ {\mathrm{MeV}}/c $}}

\newcommand{\GeV} {\mbox{$ {\mathrm{GeV}} $}}
\newcommand{\GeVc} {\mbox{$ {\mathrm{GeV}}/c $}}
\newcommand{\GeVcc} {\mbox{$ {\mathrm{GeV}}/c^2 $}}

\newcommand{\dgree} {\mbox{$ ^\circ $}}

\newcommand{\Wev} {\mbox{$ {\mathrm{W e}} \nu_{\mathrm e} $}}

\newcommand{\Zee} {\mbox{$ \Zn \ee $}}

\newcommand{\ffbar} {\mbox{$ {\mathrm f}\bar{\mathrm f} $}}
\newcommand{\qqbar} {\mbox{$ {\mathrm q}\bar{\mathrm q} $}}

\newcommand{\dm} {\mbox{$\Delta M$}}
\newcommand{\lsp}{\relax\ifmmode{\mathrm{\widetilde{\chi}^0_1}}\else${\mathrm{\widetilde{\chi}^0_1}}$\fi}

\def\PLB#1#2#3{{\rm Phys.~Lett.} {\bf{B#1}} (19#2) #3}

\def\ZPC#1#2#3{{\rm Z.~Phys.} {\bf C#1} (19#2) #3}

\def\PR#1#2#3{{\rm Phys.~Rep.} {\bf#1} (19#2) #3}

\def\NIMA#1#2#3{{\rm Nucl.~Instr.~and~Meth.} {\bf{A#1}} (19#2) #3}
\def\CPC#1#2#3{{\rm Comp.~Phys.~Comm.} {\bf#1} (19#2) #3}
\def\EJC#1#2#3{{\rm E.~Phys.~J.} {\bf{C#1}} (19#2) #3}
\newcommand{\etal}{{\it et al.}}
\def\first{\mbox{$1^{\mathrm st}$}}
\def\secnd{\mbox{$2^{\mathrm nd}$}}
\def\third{\mbox{$3^{\mathrm rd}$}}

\begin{document}
\makeatletter
\newcount\@tempcntc
\def\@citex[#1]#2{\if@filesw\immediate\write\@auxout{\string\citation{#2}}\fi
  \@tempcnta\z@\@tempcntb\m@ne\def\@citea{}\@cite{\@for\@citeb:=#2\do
    {\@ifundefined
       {b@\@citeb}{\@citeo\@tempcntb\m@ne\@citea\def\@citea{,}{\bf ?}\@warning
       {Citation `\@citeb' on page \thepage \space undefined}}%
    {\setbox\z@\hbox{\global\@tempcntc0\csname b@\@citeb\endcsname\relax}%
     \ifnum\@tempcntc=\z@ \@citeo\@tempcntb\m@ne
       \@citea\def\@citea{,}\hbox{\csname b@\@citeb\endcsname}%
     \else
      \advance\@tempcntb\@ne
      \ifnum\@tempcntb=\@tempcntc
      \else\advance\@tempcntb\m@ne\@citeo
      \@tempcnta\@tempcntc\@tempcntb\@tempcntc\fi\fi}}\@citeo}{#1}}
\def\@citeo{\ifnum\@tempcnta>\@tempcntb\else\@citea\def\@citea{,}%
  \ifnum\@tempcnta=\@tempcntb\the\@tempcnta\else
   {\advance\@tempcnta\@ne\ifnum\@tempcnta=\@tempcntb \else \def\@citea{--}\fi
    \advance\@tempcnta\m@ne\the\@tempcnta\@citea\the\@tempcntb}\fi\fi}
 
\makeatother
\begin{titlepage}
\pagenumbering{roman}
\CERNpreprint{\DpPaperGroup}{\DpPaperRef} 
\date{{\small\DpDate}} 
\title{\DpTitle} 
\address{\DpAuthors} 
\begin{shortabs} 
\noindent
%
\noindent

Searches for supersymmetric partners of top and bottom quarks are
presented using data taken by the DELPHI experiment at LEP in 1997 and 1998.
No deviations from standard model expectations are observed in
these data sets, which are taken at centre-of-mass energies of 183~GeV and
189~GeV and correspond to integrated luminosities of 54~pb$^{-1}$ and 158~pb$^{-1}$.
These results are used in combination with those obtained by
DELPHI at lower centre-of-mass energies to exclude regions
in the squark-neutralino mass plane at 95\% confidence level.
\end{shortabs}
\vfill
\begin{center}
\DpSubmit \ \\ 
\DpComment \ \\
\DpEMail \ \\
\end{center}
\vfill
\clearpage
\headsep 10.0pt
\addtolength{\textheight}{10mm}
\addtolength{\footskip}{-5mm}
\begingroup
%
\newcommand{\DpName}[2]{\hbox{#1$^{\ref{#2}}$},\hfill}
\newcommand{\DpNameTwo}[3]{\hbox{#1$^{\ref{#2},\ref{#3}}$},\hfill}
\newcommand{\DpNameThree}[4]{\hbox{#1$^{\ref{#2},\ref{#3},\ref{#4}}$},\hfill}
\newskip\Bigfill \Bigfill = 0pt plus 1000fill
\newcommand{\DpNameLast}[2]{\hbox{#1$^{\ref{#2}}$}\hspace{\Bigfill}}
%
\footnotesize
\noindent
\DpName{P.Abreu}{LIP}
\DpName{W.Adam}{VIENNA}
\DpName{T.Adye}{RAL}
\DpName{P.Adzic}{DEMOKRITOS}
\DpName{I.Ajinenko}{SERPUKHOV}
\DpName{Z.Albrecht}{KARLSRUHE}
\DpName{T.Alderweireld}{AIM}
\DpName{G.D.Alekseev}{JINR}
\DpName{R.Alemany}{VALENCIA}
\DpName{T.Allmendinger}{KARLSRUHE}
% ...
\normalsize
\endgroup
\titlefoot{Department of Physics and Astronomy, Iowa State
     University, Ames IA 50011-3160, USA
    \label{AMES}}
\titlefoot{Physics Department, Univ. Instelling Antwerpen,
     Universiteitsplein 1, B-2610 Antwerpen, Belgium \\
     \indent~~and IIHE, ULB-VUB,
     Pleinlaan 2, B-1050 Brussels, Belgium \\
     \indent~~and Facult\'e des Sciences,
     Univ. de l'Etat Mons, Av. Maistriau 19, B-7000 Mons, Belgium
    \label{AIM}}
\titlefoot{Physics Laboratory, University of Athens, Solonos Str.
     104, GR-10680 Athens, Greece
    \label{ATHENS}}
\titlefoot{Department of Physics, University of Bergen,
     All\'egaten 55, NO-5007 Bergen, Norway
    \label{BERGEN}}
\titlefoot{Dipartimento di Fisica, Universit\`a di Bologna and INFN,
     Via Irnerio 46, IT-40126 Bologna, Italy
    \label{BOLOGNA}}
\titlefoot{Centro Brasileiro de Pesquisas F\'{\i}sicas, rua Xavier Sigaud 150,
     BR-22290 Rio de Janeiro, Brazil \\
     \indent~~and Depto. de F\'{\i}sica, Pont. Univ. Cat\'olica,
     C.P. 38071 BR-22453 Rio de Janeiro, Brazil \\
     \indent~~and Inst. de F\'{\i}sica, Univ. Estadual do Rio de Janeiro,
     rua S\~{a}o Francisco Xavier 524, Rio de Janeiro, Brazil
    \label{BRASIL}}
\titlefoot{Comenius University, Faculty of Mathematics and Physics,
     Mlynska Dolina, SK-84215 Bratislava, Slovakia
    \label{BRATISLAVA}}
\titlefoot{Coll\`ege de France, Lab. de Physique Corpusculaire, IN2P3-CNRS,
     FR-75231 Paris Cedex 05, France
    \label{CDF}}
\titlefoot{CERN, CH-1211 Geneva 23, Switzerland
    \label{CERN}}
\titlefoot{Institut de Recherches Subatomiques, IN2P3 - CNRS/ULP - BP20,
     FR-67037 Strasbourg Cedex, France
    \label{CRN}}
\titlefoot{Now at DESY-Zeuthen, Platanenallee 6, D-15735 Zeuthen, Germany
    \label{DESY}}
\titlefoot{Institute of Nuclear Physics, N.C.S.R. Demokritos,
     P.O. Box 60228, GR-15310 Athens, Greece
    \label{DEMOKRITOS}}
\titlefoot{FZU, Inst. of Phys. of the C.A.S. High Energy Physics Division,
     Na Slovance 2, CZ-180 40, Praha 8, Czech Republic
    \label{FZU}}
\titlefoot{Dipartimento di Fisica, Universit\`a di Genova and INFN,
     Via Dodecaneso 33, IT-16146 Genova, Italy
    \label{GENOVA}}
\titlefoot{Institut des Sciences Nucl\'eaires, IN2P3-CNRS, Universit\'e
     de Grenoble 1, FR-38026 Grenoble Cedex, France
    \label{GRENOBLE}}
\titlefoot{Helsinki Institute of Physics, HIP,
     P.O. Box 9, FI-00014 Helsinki, Finland
    \label{HELSINKI}}
\titlefoot{Joint Institute for Nuclear Research, Dubna, Head Post
     Office, P.O. Box 79, RU-101 000 Moscow, Russian Federation
    \label{JINR}}
\titlefoot{Institut f\"ur Experimentelle Kernphysik,
     Universit\"at Karlsruhe, Postfach 6980, DE-76128 Karlsruhe,
     Germany
    \label{KARLSRUHE}}
\titlefoot{Institute of Nuclear Physics and University of Mining and Metalurgy,
     Ul. Kawiory 26a, PL-30055 Krakow, Poland
    \label{KRAKOW}}
\titlefoot{Universit\'e de Paris-Sud, Lab. de l'Acc\'el\'erateur
     Lin\'eaire, IN2P3-CNRS, B\^{a}t. 200, FR-91405 Orsay Cedex, France
    \label{LAL}}
\titlefoot{School of Physics and Chemistry, University of Lancaster,
     Lancaster LA1 4YB, UK
    \label{LANCASTER}}
\titlefoot{LIP, IST, FCUL - Av. Elias Garcia, 14-$1^{o}$,
     PT-1000 Lisboa Codex, Portugal
    \label{LIP}}
\titlefoot{Department of Physics, University of Liverpool, P.O.
     Box 147, Liverpool L69 3BX, UK
    \label{LIVERPOOL}}
\titlefoot{LPNHE, IN2P3-CNRS, Univ.~Paris VI et VII, Tour 33 (RdC),
     4 place Jussieu, FR-75252 Paris Cedex 05, France
    \label{LPNHE}}
\titlefoot{Department of Physics, University of Lund,
     S\"olvegatan 14, SE-223 63 Lund, Sweden
    \label{LUND}}
\titlefoot{Universit\'e Claude Bernard de Lyon, IPNL, IN2P3-CNRS,
     FR-69622 Villeurbanne Cedex, France
    \label{LYON}}
\titlefoot{Univ. d'Aix - Marseille II - CPP, IN2P3-CNRS,
     FR-13288 Marseille Cedex 09, France
    \label{MARSEILLE}}
\titlefoot{Dipartimento di Fisica, Universit\`a di Milano and INFN-MILANO,
     Via Celoria 16, IT-20133 Milan, Italy
    \label{MILANO}}
\titlefoot{Dipartimento di Fisica, Univ. di Milano-Bicocca and
     INFN-MILANO, Piazza delle Scienze 2, IT-20126 Milan, Italy
    \label{MILANO2}}
\titlefoot{Niels Bohr Institute, Blegdamsvej 17,
     DK-2100 Copenhagen {\O}, Denmark
    \label{NBI}}
\titlefoot{IPNP of MFF, Charles Univ., Areal MFF,
     V Holesovickach 2, CZ-180 00, Praha 8, Czech Republic
    \label{NC}}
\titlefoot{NIKHEF, Postbus 41882, NL-1009 DB
     Amsterdam, The Netherlands
    \label{NIKHEF}}
\titlefoot{National Technical University, Physics Department,
     Zografou Campus, GR-15773 Athens, Greece
    \label{NTU-ATHENS}}
\titlefoot{Physics Department, University of Oslo, Blindern,
     NO-1000 Oslo 3, Norway
    \label{OSLO}}
\titlefoot{Dpto. Fisica, Univ. Oviedo, Avda. Calvo Sotelo
     s/n, ES-33007 Oviedo, Spain
    \label{OVIEDO}}
\titlefoot{Department of Physics, University of Oxford,
     Keble Road, Oxford OX1 3RH, UK
    \label{OXFORD}}
\titlefoot{Dipartimento di Fisica, Universit\`a di Padova and
     INFN, Via Marzolo 8, IT-35131 Padua, Italy
    \label{PADOVA}}
\titlefoot{Rutherford Appleton Laboratory, Chilton, Didcot
     OX11 OQX, UK
    \label{RAL}}
\titlefoot{Dipartimento di Fisica, Universit\`a di Roma II and
     INFN, Tor Vergata, IT-00173 Rome, Italy
    \label{ROMA2}}
\titlefoot{Dipartimento di Fisica, Universit\`a di Roma III and
     INFN, Via della Vasca Navale 84, IT-00146 Rome, Italy
    \label{ROMA3}}
\titlefoot{DAPNIA/Service de Physique des Particules,
     CEA-Saclay, FR-91191 Gif-sur-Yvette Cedex, France
    \label{SACLAY}}
\titlefoot{Instituto de Fisica de Cantabria (CSIC-UC), Avda.
     los Castros s/n, ES-39006 Santander, Spain
    \label{SANTANDER}}
\titlefoot{Dipartimento di Fisica, Universit\`a degli Studi di Roma
     La Sapienza, Piazzale Aldo Moro 2, IT-00185 Rome, Italy
    \label{SAPIENZA}}
\titlefoot{Inst. for High Energy Physics, Serpukov
     P.O. Box 35, Protvino, (Moscow Region), Russian Federation
    \label{SERPUKHOV}}
\titlefoot{J. Stefan Institute, Jamova 39, SI-1000 Ljubljana, Slovenia
     and Laboratory for Astroparticle Physics,\\
     \indent~~Nova Gorica Polytechnic, Kostanjeviska 16a, SI-5000 Nova Gorica, Slovenia, \\
     \indent~~and Department of Physics, University of Ljubljana,
     SI-1000 Ljubljana, Slovenia
    \label{SLOVENIJA}}
\titlefoot{Fysikum, Stockholm University,
     Box 6730, SE-113 85 Stockholm, Sweden
    \label{STOCKHOLM}}
\titlefoot{Dipartimento di Fisica Sperimentale, Universit\`a di
     Torino and INFN, Via P. Giuria 1, IT-10125 Turin, Italy
    \label{TORINO}}
\titlefoot{Dipartimento di Fisica, Universit\`a di Trieste and
     INFN, Via A. Valerio 2, IT-34127 Trieste, Italy \\
     \indent~~and Istituto di Fisica, Universit\`a di Udine,
     IT-33100 Udine, Italy
    \label{TU}}
\titlefoot{Univ. Federal do Rio de Janeiro, C.P. 68528
     Cidade Univ., Ilha do Fund\~ao
     BR-21945-970 Rio de Janeiro, Brazil
    \label{UFRJ}}
\titlefoot{Department of Radiation Sciences, University of
     Uppsala, P.O. Box 535, SE-751 21 Uppsala, Sweden
    \label{UPPSALA}}
\titlefoot{IFIC, Valencia-CSIC, and D.F.A.M.N., U. de Valencia,
     Avda. Dr. Moliner 50, ES-46100 Burjassot (Valencia), Spain
    \label{VALENCIA}}
\titlefoot{Institut f\"ur Hochenergiephysik, \"Osterr. Akad.
     d. Wissensch., Nikolsdorfergasse 18, AT-1050 Vienna, Austria
    \label{VIENNA}}
\titlefoot{Inst. Nuclear Studies and University of Warsaw, Ul.
     Hoza 69, PL-00681 Warsaw, Poland
    \label{WARSZAWA}}
\titlefoot{Fachbereich Physik, University of Wuppertal, Postfach
     100 127, DE-42097 Wuppertal, Germany
    \label{WUPPERTAL}}
\addtolength{\textheight}{-10mm}
\addtolength{\footskip}{5mm}
\clearpage
\headsep 30.0pt
\end{titlepage}
%
\pagenumbering{arabic} 
\setcounter{footnote}{0} %
\large
%
\section{Introduction \label{sec:INTRO}}

This paper reports on a search for scalar partners of 
quarks (squarks) in data taken by DELPHI in 1997 and 1998 at
centre-of-mass energies ($\sqrt{s}$) of 183~GeV and 189~GeV. 
Mass limits for these particles have already been published based on
data taken at LEP2~\cite{DELLIM},~\cite{LEPLIM}.

Scalar partners of right- and left-handed fermions are predicted by
supersymmetric models and, in particular, by the minimal
supersymmetric extension of the standard model (MSSM) \cite{MSSM}.
They could be produced pairwise via $e^+e^-$ annihilation into
\Zn/$\gamma$. 
Large Yukawa coupling running for the 
diagonal elements and important off-diagonal terms make the partners of heavy
fermions as the most probable candidates for the charged lightest
supersymmetric particle.
As a consequence their lighter states are candidates for the
lightest charged supersymmetric particle.

Throughout this paper conservation of R-parity is assumed, which
implies that the lightest supersymmetric particle (LSP) is stable.
The LSP is assumed to be the lightest neutralino which
interacts only weakly 
with matter, such that events will be characterised by missing
momentum and energy.  

In a large fraction of the MSSM parameter space sfermions are predicted to
decay dominantly into the corresponding fermion and the lightest
neutralino. 
Consequently in the search for sbottom particles only the 
decay into $b + {\tilde {\chi}}^o_1$ was considered.
For the stop squark, the equivalent decay into 
$t + {\tilde {\chi}}^o_1$ is kinematically not allowed at LEP,
and the decay of a stop into a bottom quark and a
chargino is disfavoured in view of existing limits on the chargino
mass \cite{CHLIM}.  The dominant two-body decay channel is thus
the one into a charm quark and a neutralino. 

\section{Detector description \label{sec:DETEC}}

The DELPHI detector and its performance have been described in detail 
elsewhere~\cite{DELPHIPER1,DELPHIPER2}; only those components relevant
for the present analyses are discussed here. 
Charged particle tracks are reconstructed in the 1.2 T solenoidal 
magnetic field by a system of cylindrical tracking chambers.
These are the Vertex Detector (VD),
the Inner Detector (ID), the Time Projection Chamber
(TPC) and the Outer Detector (OD). In addition,
two planes of drift chambers aligned
perpendicular to the  beam axis (Forward Chambers A and B) track
particles in the forward and backward directions, covering polar
angles $11\dgree<\theta<33\dgree$ and $147\dgree<\theta<169\dgree$
with respect to the beam ($z$) direction.

The VD consists of three cylindrical layers of 
silicon detectors, at radii 6.3~cm, 9.0~cm and 11.0~cm.
All three layers measure 
coordinates in the plane transverse to the beam.
The inner (6.3~cm) and the outer (11.0~cm) layers
contain double-sided detectors to also measure $z$ coordinates.
The VD covers polar angles from
$24\dgree$ to $156\dgree$.
The ID consists of
a cylindrical drift chamber (inner radius 12
cm and outer radius 22 cm) covering polar angles between $15\dgree$
and $165\dgree$. 
The TPC, the principal tracking device of 
DELPHI, consists of a cylinder of 30 cm inner
radius, 122 cm outer radius and has a length of 2.7 m. Each 
end-plate has been divided 
into 6 sectors, with 192 sense wires used for the dE/dx measurement
and 16 circular pad rows  used for 3 dimensional 
space-point reconstruction.
The OD consists of
5 layers of drift cells at radii
between 192~cm and 208~cm, covering polar angles between $43\dgree$
and $137\dgree$.

The average momentum resolution for the charged particles in hadronic final
states is in the range 
$\Delta p/p^2 \simeq 0.001$ to 
$0.01 (\GeVc)^{-1}$, depending on which detectors
are used in the track fit \cite{DELPHIPER2}.

The electromagnetic calorimeters consist of the High density Projection
Chamber~(HPC) covering the barrel region of
$40\dgree<\theta<140\dgree$, the Forward ElectroMagnetic Calorimeter~(FEMC) 
covering $11\dgree<\theta<36\dgree$ and
$144\dgree<\theta<169\dgree$, and the STIC, a Scintillator TIle
Calorimeter which extends the coverage down to 1.66$\dgree$ from the beam
axis in both directions. The 40$\dgree$ taggers are made of
single layer scintillator-lead counters used to veto electromagnetic
particles that may be not measured in the region between
the HPC and FEMC.  The efficiency to register a photon with energy
above 5~GeV at polar angles between $20\dgree$ and $160\dgree$, measured with
the LEP1 data, 
is greater than 99$\%$~\cite{DELPHIPER2}. The hadron
calorimeter (HCAL) covers 98$\%$ of the solid angle. Muons with
momenta above 2 GeV/c penetrate the HCAL and are recorded in a set of
Muon Drift Chambers.

Decays of b-quarks are tagged using a probabilistic method based on
the impact parameters of tracks with respect to the main vertex. 
${\mathcal P}^+_E$ stands for the
corresponding probability estimator for tracks with positive impact parameters,
the sign of the impact parameter being defined by the jet direction.
The combined probability estimator ${\mathcal P}_{com b}$ includes in addition
contributions from properties of reconstructed secondary
vertices~\cite{comb}.
\section{Data samples and event generators \label{sec:SAMPLES}}

Data were taken during the 1997 and 1998 LEP runs at mean centre-of-mass
energies of 183~GeV and 
189~GeV, corresponding to integrated luminosities of 
54~pb$^{-1}$ and 158~pb$^{-1}$.

Simulated events were generated with several programs in
order 
to evaluate the signal efficiency and the background contamination.
All the models used
{\tt JETSET}~7.4~\cite{JETSET} for quark fragmentation with 
parameters tuned to represent DELPHI data~\cite{DELPHIDATA}.

Stop events were generated according to
the expected differential
cross-sections, using the {\tt BASES} and {\tt SPRING}
program packages~\cite{SPRING}.
Special care was
taken in the modelling of the stop hadronisation~\cite{BEENAKKER}.
Sbottom events were generated with the {\tt SUSYGEN} program~\cite{SUSYGEN}.
The background processes \eeto\ \qqbar ($n\gamma$) and processes
leading to four-fermion final states, $(\Zn/\gamma)^*(\Zn/\gamma)^*$, 
$\Wp ^*\Wm ^*$, \Wev, and \Zee\ were 
generated using {\tt PYTHIA}~\cite{JETSET}.
At the generator level, the cut on the invariant mass
of the virtual $(\Zn/\gamma)^*$ in  
the $(\Zn/\gamma)^*(\Zn/\gamma)^*$ process 
was set at  2~\GeVcc , in order to be able
to estimate the background from low mass \ffbar\ pairs.
The calculation of the four-fermion 
background was cross-checked using the program 
{\tt EXCALIBUR}~\cite{EXCALIBUR}, which consistently
takes into account all amplitudes leading to a given four-fermion
final state. The version of {\tt EXCALIBUR} used does not, 
however, include the transverse
momentum of initial state radiation.
Two-photon interactions leading to hadronic final states
were simulated using {\tt TWOGAM}~\cite{TWOGAM} and {\tt BDKRC}~\cite{BDK}
for the Quark Parton Model contribution. Leptonic final states
with muons and taus were also modelled with {\tt BDKRC}. {\tt BDK}~\cite{BDK}
was used for final states with electrons only.

Generated signal and background events were passed through detailed
detector response simulation~\cite{DELPHIPER2} and processed with the
same reconstruction and analysis programs as the real data.
The number of background events simulated is 
mostly several times larger than the number expected in the real data.

\section{Event selection \label{sec:SQUARK}}

In this section the selection to search for stop and sbottom in the decay modes
c\XN{1}\: and b\XN{1}\:, respectively, is presented. In both cases the
experimental signatures consist of events with two jets and missing
momentum.  Since event parameters, such as visible energy, greatly depend
on the mass difference \dm\: between the squark and the LSP, optimised
selection procedures are used for the degenerate ($\dm \leq
10\:\GeVcc$), and the non-degenerate ($\dm > 10\:\GeVcc$) mass case.
The main differences between stop and sbottom events arise from the
hadronisation, which occurs either before (\stq) or after (\sbq) the
decay of the scalar quark (in a large fraction of the MSSM parameter
space the width of the sbottom decay into $b + {\tilde {\chi}}^o_1$ 
is greater than the typical QCD scale so that the sbottom does not hadronize
before it decays).
These differences are visible in particular
in the degenerate mass case. Consequently different selections are
used for the stop and sbottom analyses in the degenerate mass case
whereas the selections are identical in the non-degenerate mass case.

In a first step particles are selected and clustered into jets
using the Durham algorithm~\cite{DURHAM} with $y_{cut}=0.08$.
Reconstructed charged particles are required to have momenta above
100~\MeVc\: and impact parameters to the measured interaction point
below 4~cm in the transverse plane
and below 10~cm in the beam direction. Clusters in the calorimeters
are interpreted as neutral particles if they are not associated to
charged particles, and if their energy exceeds 100~\MeV.

In the second step of the analysis, hadronic events are selected.
Only two-jet events are accepted. The following requirements are
optimised separately for the two \dm\: regions:
\begin{description}[3em]
\item[{\bf Non-degenerate mass case}:] 
  For both the stop and sbottom
  analyses hadronic events are selected by requiring at least eight
  charged particles, a total transverse energy\footnote{The transverse
  energy $E_t$ of a particle is defined as $E_t = \sqrt { E^2_x + E^2_y}$ where
  $E_x$ and $E_y$ respectively are $E {\cos {\phi}} {\sin {\theta}}$ and
  $E {\sin {\phi}} {\sin {\theta}}$. The angles $\phi$ and $\theta$ are 
  respectively the azimuthal and polar angle of the particle.}    
  greater than 15~\GeV\:
  and a transverse energy of the most energetic jet greater than
  10~\GeV.  These three cuts are aimed at reducing the
  background coming from two-photon processes. Forward Bhabha
  scattering is suppressed by requiring that the total energy in the
  FEMC is lower than 25~GeV.  Z$^0$($\gamma$) processes with a detected
  photon are reduced by requiring that the total energy in the HPC
  is lower than 40~GeV. 
  Finally, at~$\sqrt{s}$~=~183~GeV, the requirement for
  substantial missing energy is fulfilled by demanding that the quantity
  $\sqrt {s^{\prime}}$ is lower than 170~GeV. The quantity $\sqrt {s^{\prime}}$
  is the effective centre-of-mass energy after photons radiation from
  the incoming $e^+e^-$ beams.
  At~$\sqrt{s}$~=~189~GeV,
  this requirement is replaced by the requirement that
  the polar angles of the two jets are between
  $20\dgree$ and $160\dgree$. 
\item[{\bf Degenerate mass case}:] To select hadronic events in 
  the stop analysis the number of charged particles is required to be
  greater than five, the total charged energy has to be lower than
  $0.3 \sqrt s$ (in order to select events with missing energy) and the
  polar angle of the total missing momentum has to be between
  $15\dgree$ and $165\dgree$, in order to reduce
  the background from radiative return events. The total energy in
  the FEMC and HPC has to be lower than 10~\GeV\: and 40~\GeV,
  respectively. The reduction of two-photon processes is ensured by
  requiring that the total transverse energy is greater than 5~\GeV\:
  and that the quantity $p_{tt} = \sqrt {p_{tt1}^2 + p_{tt2}^2} $ is
  greater than 5~\GeVc, where $p_{tti}$ is the transverse momentum of
  jet $i$ with respect to the thrust axis projected onto the plane
  transverse to the beam axis.
  Finally, the most energetic charged particle is required to have a
  polar angle between $30\dgree$ and $150\dgree$ and a momentum greater
  than 2~\GeVc. Similarly the polar angle of the most energetic
  neutral particle is required to be between $20\dgree$ and $160\dgree$.
  At~$\sqrt{s}$~=~183~GeV, the
  sbottom selection at this step is similar to the stop analysis described
  above except for the requirement on $p_{tt}$ which is replaced by 
  requiring the ratio $p_{tt}/E_{tot}$ to be greater than 50\% where $E_{tot}$
  is the total energy of the event.
  At~$\sqrt{s}$~=~189~GeV, the sbottom selection at this step is simplified
  by removing the above requirement on $p_{tt}/E_{tot}$.   
\end{description}

\begin{figure}[htbp]
\begin{center}
\begin{tabular}{c}
\epsfig{file=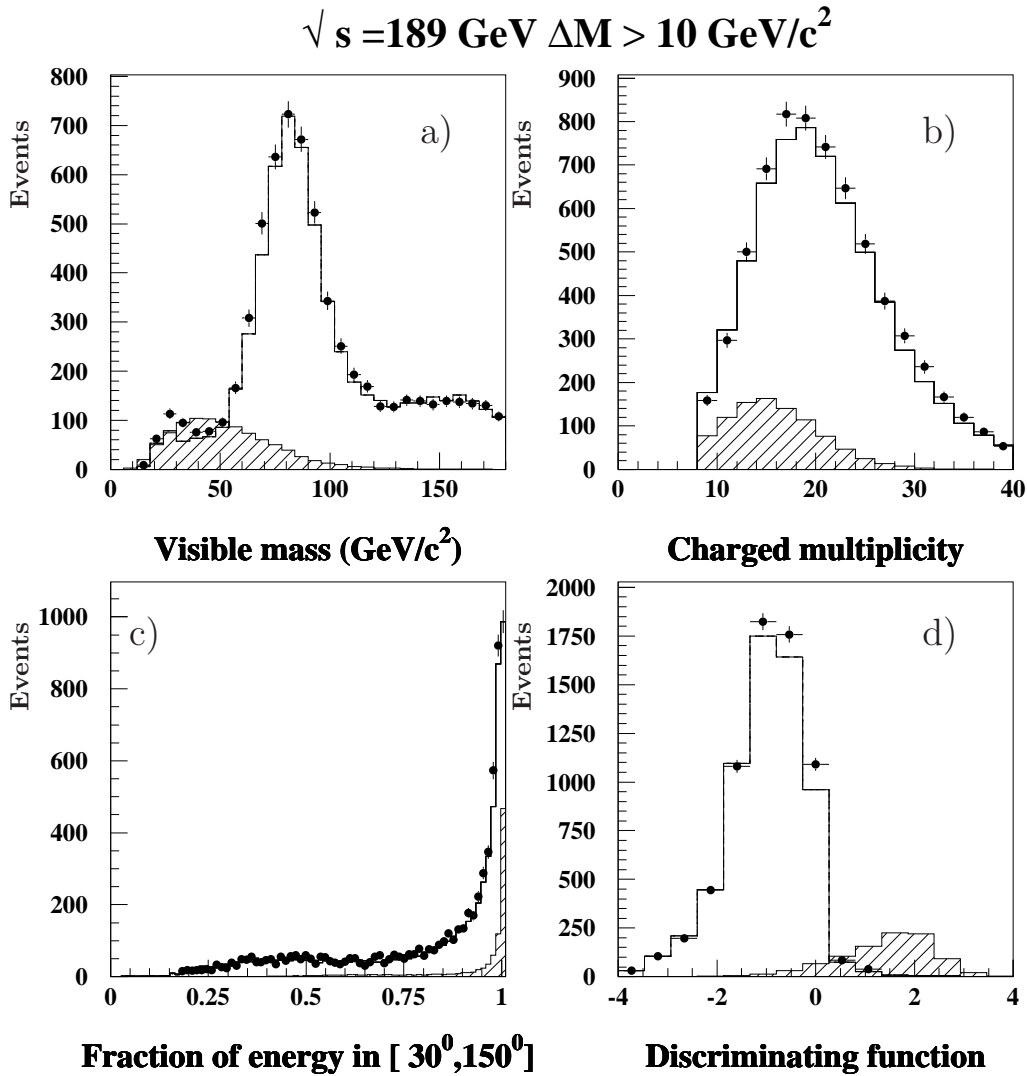,width=15cm} \\
\end{tabular}
{\Large
\put (-270.,150.){a)}
\put (-80.,150.){b)}
\put (-380.,-40.){c)}
\put (-80.,-40.){d)}
}
{\normalsize
\put (-425.,125.){\rotatebox{90}{\bf Events}}
\put (-235.,125.){\rotatebox{90}{\bf Events}}
\put (-425.,-65.){\rotatebox{90}{\bf Events}}
\put (-235.,-65.){\rotatebox{90}{\bf Events}}
}
\caption{\small a) the visible mass, b) the charged multiplicity, c) the fraction
of the energy in the polar angle interval $[30^o,150^o]$ and d) the discriminating
function against the $Z \gamma$ background (as described in the text)
for the non-degenerate mass case concerning both stop and sbottom analysis.
The dots with error bars show the data while the clear histogram is the SM prediction.
Each hatched area shows the stop signal for stop masses of
70~\GeVcc\:, 80~\GeVcc\: and 90~\GeVcc\: with $\Delta M >$ 10~\GeVcc\: 
(with a normalization factor to the luminosity in the range 8 to 90) 
where $\Delta M$ represents
the mass difference between the squark and the LSP.
The case $\dm > 10\:\GeVcc$ corresponds to the non-degenerate mass case.}
\label{fig1}
\end{center}
\end{figure}

\begin{figure}[htbp]
\begin{center}
\begin{tabular}{c}
\epsfig{file=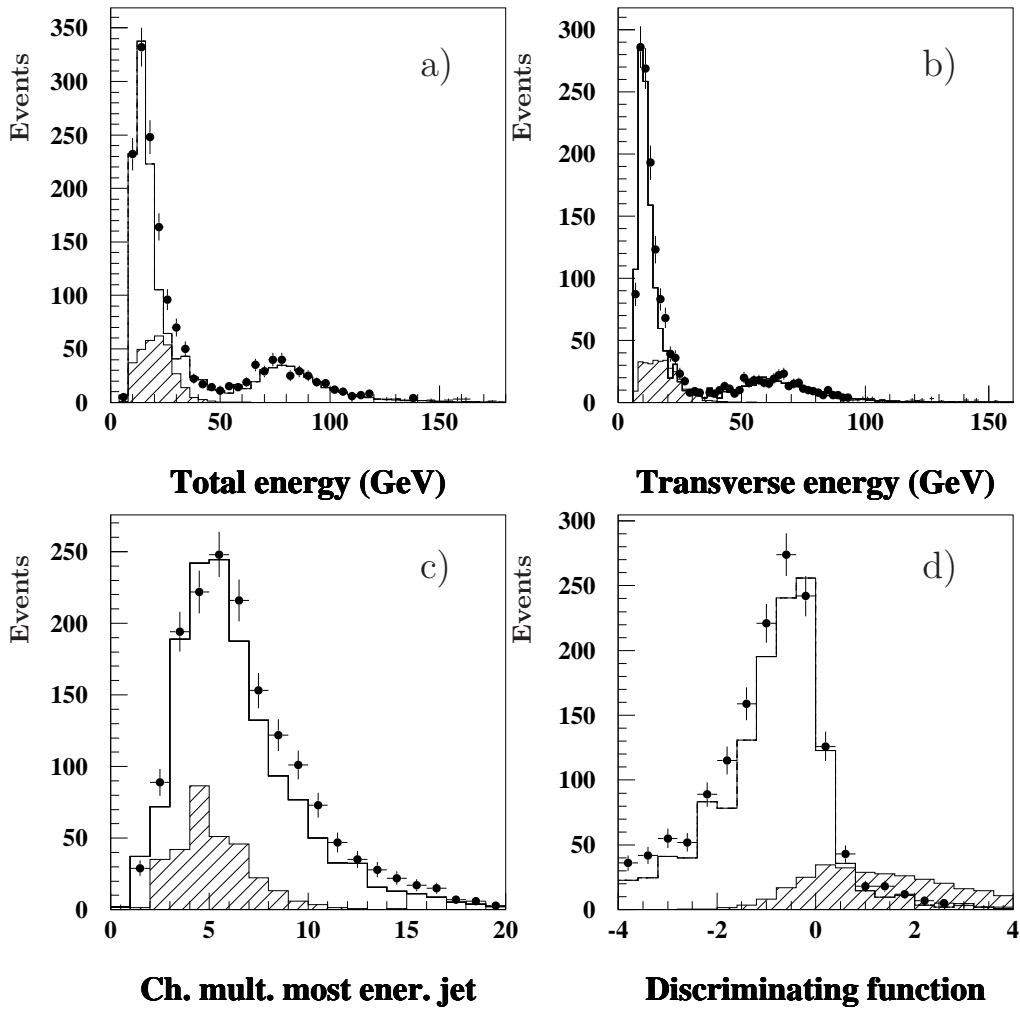,width=15cm} \\
\end{tabular}
{\Large
\put (-270.,150.){a)}
\put (-80.,150.){b)}
\put (-270.,-40.){c)}
\put (-80.,-40.){d)}
}
{\normalsize
\put (-425.,125.){\rotatebox{90}{\bf Events}}
\put (-235.,125.){\rotatebox{90}{\bf Events}}
\put (-425.,-65.){\rotatebox{90}{\bf Events}}
\put (-235.,-65.){\rotatebox{90}{\bf Events}}
}
\caption{\small a) the total energy, b) the transverse energy,
c) the charged multiplicity of the leading jet 
and d) the discriminating function (as described 
in the text)
for the degenerate mass case of the stop analysis. 
The dots with error bars show the data while the clear histogram is the SM prediction.
Each hatched area shows the stop signal for stop masses of
70~\GeVcc\:, 80~\GeVcc\: and 90~\GeVcc\: with $\Delta M \leq$ 10~\GeVcc\: 
(with a normalization factor to the luminosity in the range 8 to 90) 
where $\Delta M$ represents
the mass difference between the squark and the LSP.
The case $ \dm \leq 10\:\GeVcc $ corresponds to the degenerate mass case.}
\label{fig2}
\end{center}
\end{figure}

\begin{figure}[htb]
\begin{center}
\begin{tabular}{c}
\epsfig{file=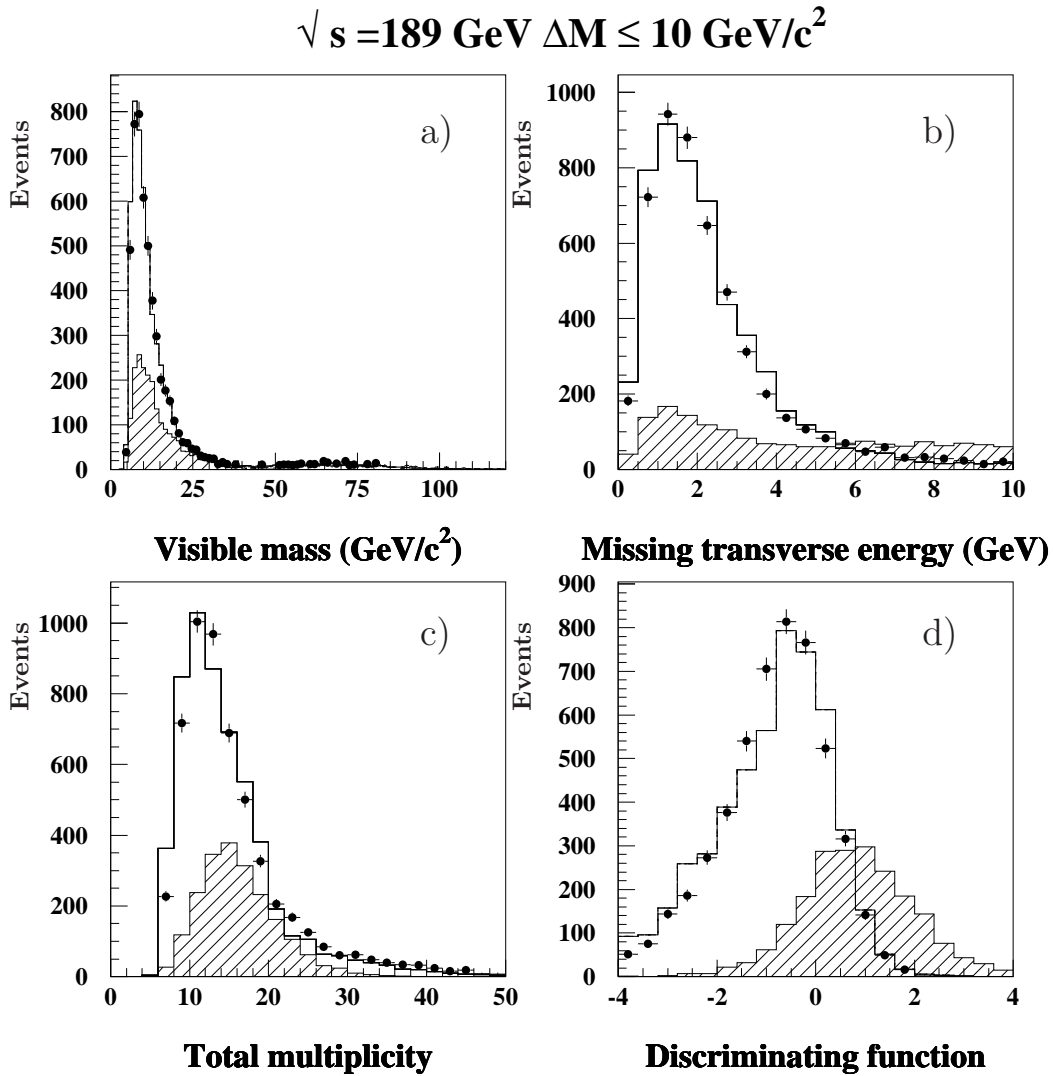,width=15cm} \\
\end{tabular}
{\Large
\put (-270.,150.){a)}
\put (-80.,150.){b)}
\put (-270.,-40.){c)}
\put (-80.,-40.){d)}
}
{\normalsize
\put (-425.,125.){\rotatebox{90}{\bf Events}}
\put (-235.,125.){\rotatebox{90}{\bf Events}}
\put (-425.,-65.){\rotatebox{90}{\bf Events}}
\put (-235.,-65.){\rotatebox{90}{\bf Events}}
}
\caption{\small a) the visible mass, b) the missing transverse energy,
c) the total multiplicity and d) the discriminating function
(as described in the text) 
for the degenerate mass case of the sbottom analysis.
Each hatched area shows the sbottom signal for sbottom masses
of 50~\GeVcc\:, 60~\GeVcc\:, 70~\GeVcc\:, 80~\GeVcc\: 
and 90~\GeVcc\: with $\Delta M \leq$ 10~\GeVcc\: 
(with a normalization factor to the luminosity in the range 5 to 100) 
where 
$\Delta M$ represents
the mass difference between the squark and the LSP.
The case $ \dm \leq 10\:\GeVcc $ corresponds to the degenerate mass case.}
\label{fig3}
\end{center}
\end{figure}

After this second step and 
for both the non-degenerate and the degenerate mass cases, agreement between
data and expectations from the Monte Carlo simulation describing standard 
model processes
is found to be good as can be seen from Figures~\ref{fig1}a-c
showing the visible mass, the charged multiplicity and the fraction
of the energy for polar angles between $30^o$ and $150^o$ at $\sqrt{s}$ = 189~GeV.
Figures~\ref{fig2}a-c show the total energy, the transverse energy and
the charged multiplicity of the leading jet,
for the degenerate mass case of the stop analysis at $\sqrt{s}$ = 189~GeV. 
Figures~\ref{fig3}a-c show
the visible mass, the missing transverse energy and
the total multiplicity for the degenerate mass case of
the sbottom analysis at $\sqrt{s}$ = 189~GeV.

In a third step discriminating linear functions~\cite{discri} are
used in order to achieve optimum rejection power.
They have been
determined in the following way:
\begin{description}[3em]
\item[{\bf Non-degenerate mass case}:]
  In this case, the same functions have
  been used both for the stop and the sbottom analysis. A first
  discriminating linear function has been determined using training
  samples of signal and Z$^0$($\gamma$) background processes. 
  For the training of a second discriminating linear function, signal
  and WW background event samples have been used. In the
  non-degenerate mass case, these two sources of background processes
  are found to be dominant after the first and second step of the
  event selection. 

\item[{\bf Degenerate mass case}:] Here the main source of background
  remaining after the first and second step of the event selection is
  found to be $\gamma\gamma$ events. Different functions have been
  determined for the stop and sbottom analyses using training samples
  of signal and two-photon events.
\end{description}

Figure~\ref{fig1}d shows the discriminating
function against the Z$^0 \gamma$ background 
for the non-degenerate mass case at $\sqrt{s}$~=~189~GeV, Figure~\ref{fig2}d 
and~\ref{fig3}d show the discriminating functions for the degenerate
mass domains of the stop and sbottom analyses at $\sqrt{s}$~=~189~GeV.
For these degenerate and non-degenerate mass cases, fair agreements between
data and expectations from Monte Carlo describing standard model processes
are found. The data and Monte Carlo small disagreement 
of the discriminating function for the degenerate mass cases shown in
Figure~\ref{fig2}d is restricted to the negative values of this function which
correspond to the region of the bulk of the expectations from Monte Carlo describing
standard model processes in particular two-photon interactions leading to hadronic
final states which are known to be difficult to modelize. This region
does not correspond to the squark signal region. As shown by the hatched areas
of Figure~\ref{fig2}d,
the positive values of this
discriminating function correspond to the squark signal region and in this
region the agreement between data and expectations from Monte Carlo describing
standard model processes is very good.
\par
The final background reduction is performed by sequential cuts. In
the non-degenerate mass case, one set of cuts is used to select both
stop and sbottom events. It is shown, together with the number of
events retained in data and background simulation, 
in~Tables~\ref{tab:stsb:sel4-183}~and~\ref{tab:stsb:sel4}. 
\par
In the degenerate mass case two different
selections are used for stop and sbottom.
These are shown in
Table~\ref{tab:stop:sel4} for the stop analysis 
at~$\sqrt{s}$~=~183~GeV~and~$\sqrt{s}$~=~189~GeV and in~Tables~\ref{tab:sbot:sel4-183}
and~\ref{tab:sbot:sel4}
for the sbottom analysis at~$\sqrt{s}$~=~183~GeV
and~$\sqrt{s}$~=~189~GeV respectively.
\section{Results \label{sec:RESU}}
\begin{figure}[htb]
\begin{center}
\epsfig{file=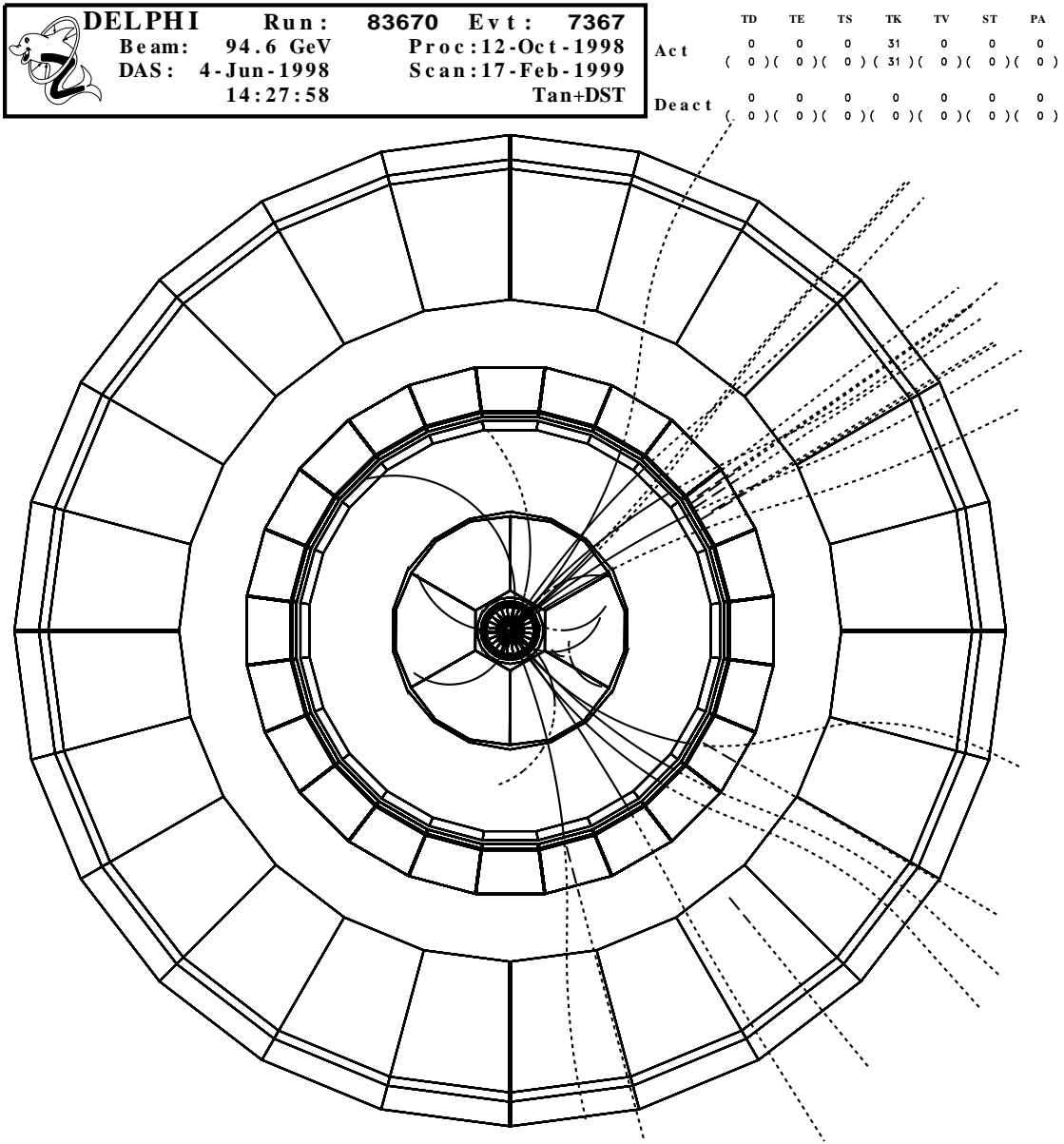,width=15cm} 
\caption[]{\small View of one candidate event from the non-degenerate mass case
in the transverse plane. The corresponding total energy is 57.3 GeV,
the charged multiplicity is found to be 27, the total visible mass is 43.3 $\GeVcc$,
the polar angle of the missing momentum is 74.8 degrees and the polar angle 
of the two
jets are 86.5 degrees and 125 degrees respectively.}
\label{fig0}
\end{center}
\end{figure}

\begin{figure}[htb]
\begin{center}
\begin{tabular}{c}
\epsfig{file=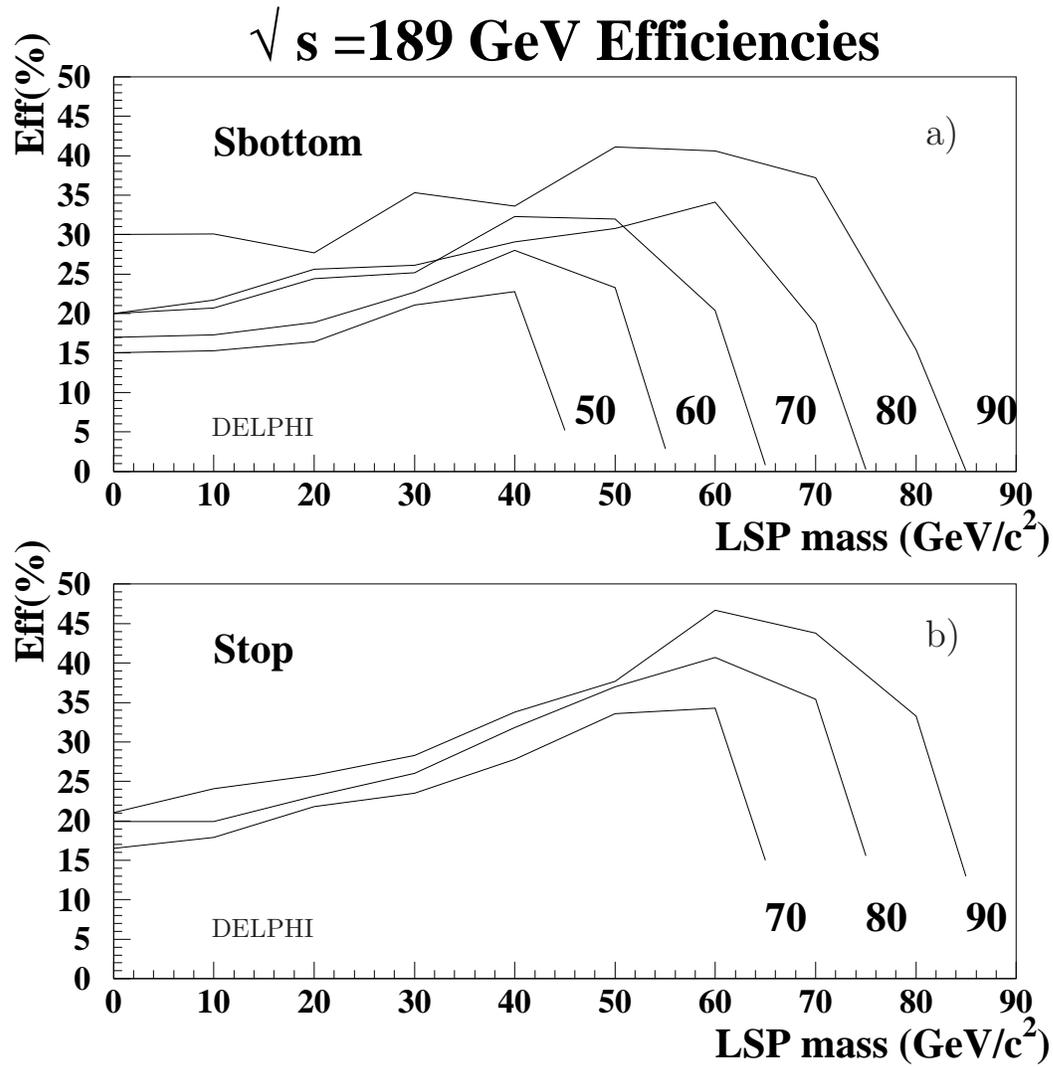,width=15cm} \\
\end{tabular}
{\Large
\put (-80.,150.){a)}
\put (-80.,-40.){b)}
}
{\normalsize
\put (-350.,40.){DELPHI}
\put (-350.,-150.){DELPHI}
}
\caption{\small Efficiencies for the a) sbottom and b) stop selection
in the search for two-body decays as function of the LSP mass
for various sbottom and stop masses. The sbottom and stop masses 
are indicated on the
plots in units of GeV/$c^2$.}
\label{fig:stsb:eff}
\end{center}
\end{figure}

The number of candidates found and the expected background levels are
shown in Table~\ref{tab:stsb:bkg-183} and~Table~\ref{tab:stsb:bkg}. 
There are candidates
in common in
the stop and sbottom analyses. The total background is given
assuming a $\prime \prime$or$\prime \prime$ between the degenerate mass case and the
non-degenerate mass case for
the stop and sbottom analyses.
One candidate event from the non-degenerate mass case analysis
is shown in Figure~\ref{fig0}.
The efficiencies of the stop and
sbottom signal selection are summarised in Figure~\ref{fig:stsb:eff}.  They
have been evaluated using 35 simulated samples at different points in
the ($M_{\tilde{\mathrm q}}$,\MXN{1}) plane, for squark masses between
50 and 90~\GeVcc\: and neutralino masses between 0 and 85~\GeVcc.
\par
\begin{figure}[htb]
\begin{center}
\begin{tabular}{c}
\epsfig{file=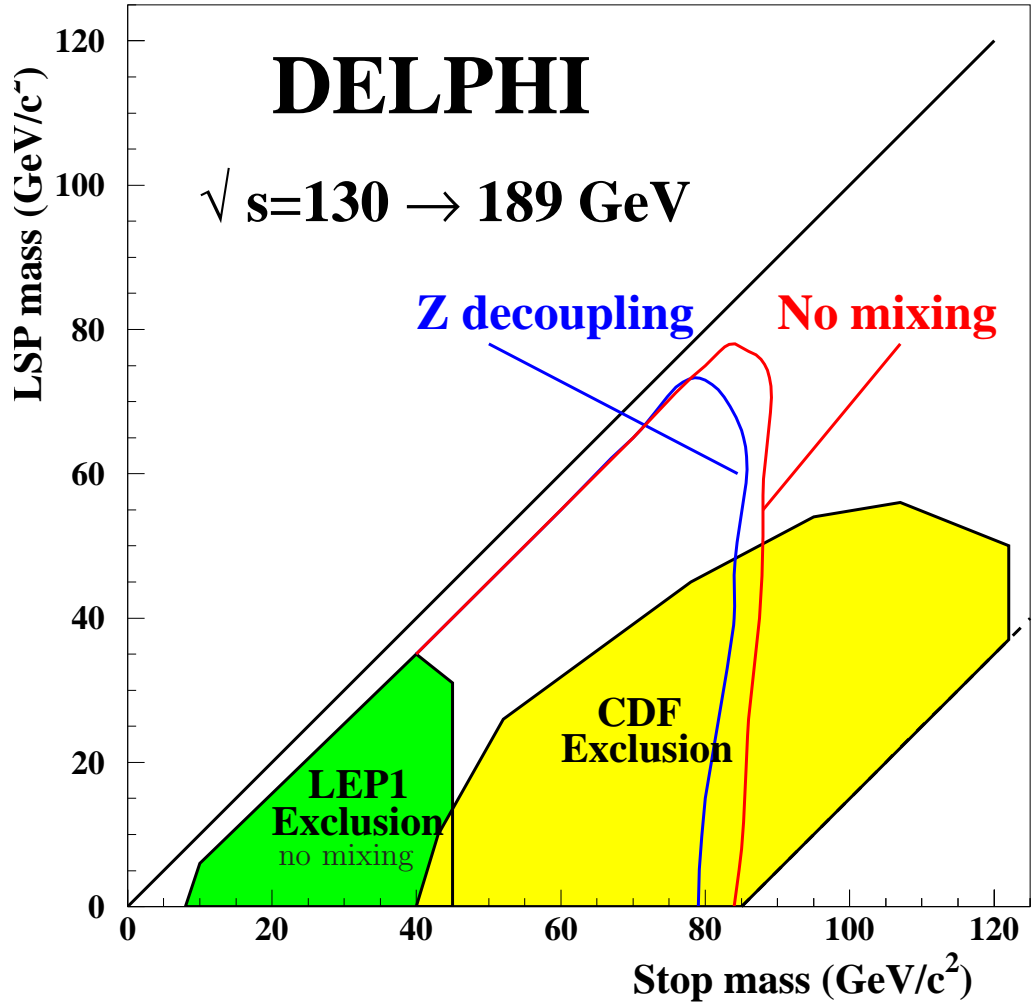,width=15cm} 
\end{tabular}
\put (-330.,-150.){no mixing}
\caption[]{\small Exclusion domains at 95\% confidence level 
in the (${\tilde t}$,${\tilde {\chi}}^o_1$) mass plane assuming
100\% branching ratio into c${\tilde {\chi}}^o_1$ for pure 
left-handed state ($\theta = 0$ rad) and for the minimum cross-section
($\theta = 0.98 $ rad) corresponding to the decoupling of the stop from the
Z boson.
The limits are obtained combining data at $\sqrt s$ = 130 - 189~GeV. 
The shaded areas have been excluded by~LEP1~\cite{stoplep1}~and~CDF~\cite{stopd0}.}
\label{lim1}
\end{center}
\end{figure}
\begin{figure}[htb]
\begin{center}
\begin{tabular}{c}
\epsfig{file=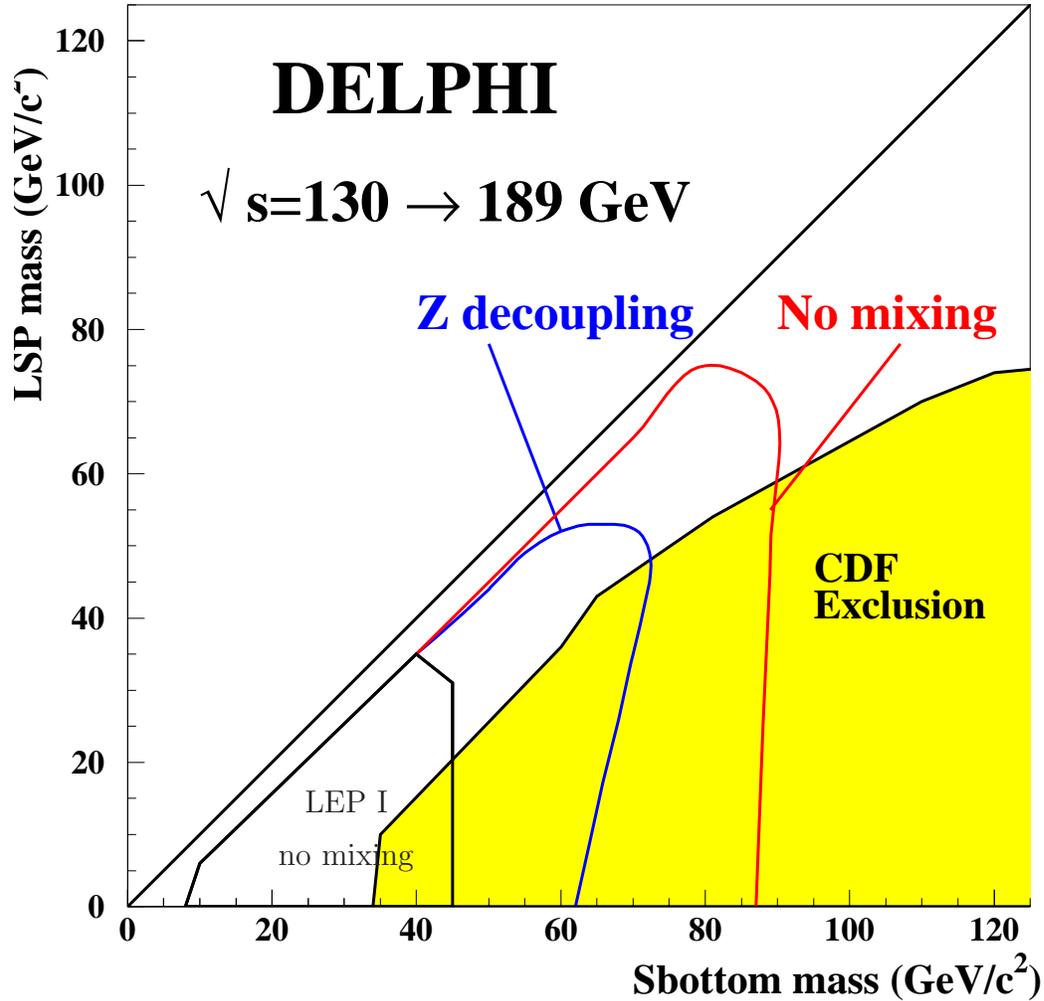,width=15cm} \\
\end{tabular}
\put (-320.,-130.){LEP I}
\put (-330.,-150.){no mixing}
\caption[]{\small Exclusion domains at 95\% confidence level 
in the (${\tilde b}$,${\tilde {\chi}}^o_1$) mass plane assuming
100\% branching ratio into b${\tilde {\chi}}^o_1$ for pure 
left-handed state ($\theta = 0$ rad) and for the minimum cross-section
($\theta = 1.17$ rad) corresponding to the decoupling of the sbottom from the
Z boson.
The limits are obtained combining data at $\sqrt s$ = 130 - 189 GeV.}
\label{lim2}
\end{center}
\end{figure}

No evidence for stop or sbottom production has been found in 
the two-body decay channels.
Figure~\ref{lim1} and Figure~\ref{lim2} show 
the ($M_{\tilde{\mathrm q}}$,\MXN{1})
regions excluded at 95\% confidence level by the search for
$\stq \to {\mathrm c}\XN{1}$  and 
$\sbq \to {\mathrm b}\XN{1}$  decays,
with the 100\% branching ratio assumption,
both for purely left-handed states (with maximum cross-section)
and the states with minimum cross-section.
We have also used the results 
(efficiencies, number of candidates and expected
background) of the analyses of the data at 130 - 172~GeV~\cite{DELLIM}~in
order to derive these exclusion regions.
\par
In order to estimate systematic errors coming from detector 
effects and modelling, the differences of 
the mean values of the observables used for the above analyses 
(sequential cuts steps and discriminating linear analyses steps)  
between real data and simulation are calculated at the level of
the first step of the selection described in section~\ref{sec:SQUARK}.
The difference $\delta$ for the mean value of the observable $X$ between
real data and simulation is used in order to shift $X$ 
according to $X + \delta$ and $X - \delta$. The analyses described
in section~\ref{sec:SQUARK} then use the shifted observables and
the differences in efficiencies and expected background with respect
to efficiencies and expected background obtained with the unshifted
observables
are taken as systematic errors.
The relative systematic errors for efficiencies are 10\%
in the non-degenerate mass case and 15\% in the
degenerate mass case. 
The systematic errors for the expected background are given
in Table~\ref{tab:stsb:bkg-183} and~Table~\ref{tab:stsb:bkg}.
\par
Systematic errors on efficiencies coming from the modelling of the
hadronization of the stop are estimated by switching off the
hadronization of the stop. The relative systematic errors for efficiencies are 
2\% in the non degenerate mass case and 8\% in the
degenerate mass case.

\section{Conclusions \label{sec:CONCLU}}

In data samples of 54 pb$^{-1}$ and 158 pb$^{-1}$ collected by the DELPHI
detector at centre-of-mass energies of 183~GeV and 189~GeV searches are
performed for events with acoplanar jet pairs. 
The results are
combined with those already obtained 
at centre-of-mass energies between $130$--$172$~\GeV\:.

At 183~GeV, the search for stop and sbottom quarks,
decaying into c\XN{1}\: and b\XN{1}, respectively, 
gives in total 3 candidates (some 
candidates are 
in common in
the stop and sbottom analyses)
 well
compatible with the expected background 
of $3.4 \pm  0.5 $. 

At 189~GeV, the search for stop and sbottom quarks,
decaying into c\XN{1}\: and b\XN{1}, respectively, 
gives in total 9 candidates (some 
are candidates are also
in common in
the stop and sbottom analyses)
 well
compatible with the expected background 
of $11.6 \pm  1.4$. 

For the stop, a mass limit of
79~\GeVcc\: is obtained for the state with minimal cross-section, 
if the mass difference 
between the squark and the LSP is above 15~\GeVcc\:. A mass limit 
of 62~\GeVcc\: is obtained for the sbottom quark under the same 
condition. 

In the case of maximum cross-section, these numbers are 84~\GeVcc\: for
the stop and  87~\GeVcc\: for the sbottom.

\section*{Acknowledgements}
\vskip 3 mm We are greatly indebted to our technical 
collaborators, to the members of the CERN-SL Division for the excellent 
performance of the LEP collider, and to the funding agencies for their
support in building and operating the DELPHI detector.\\
We acknowledge in particular the support of \\
Austrian Federal Ministry of Science and Traffics, GZ 616.364/2-III/2a/98, \\
FNRS--FWO, Belgium,  \\
FINEP, CNPq, CAPES, FUJB and FAPERJ, Brazil, \\
Czech Ministry of Industry and Trade, GA CR 202/96/0450 and GA AVCR A1010521,\\
Danish Natural Research Council, \\
Commission of the European Communities (DG XII), \\
Direction des Sciences de la Mati$\grave{\mbox{\rm e}}$re, CEA, France, \\
Bundesministerium f$\ddot{\mbox{\rm u}}$r Bildung, Wissenschaft, Forschung 
und Technologie, Germany,\\
General Secretariat for Research and Technology, Greece, \\
National Science Foundation (NWO) and Foundation for Research on Matter (FOM),
The Netherlands, \\
Norwegian Research Council,  \\
State Committee for Scientific Research, Poland, 2P03B06015, 2P03B1116 and
SPUB/P03/178/98, \\
JNICT--Junta Nacional de Investiga\c{c}\~{a}o Cient\'{\i}fica 
e Tecnol$\acute{\mbox{\rm o}}$gica, Portugal, \\
Vedecka grantova agentura MS SR, Slovakia, Nr. 95/5195/134, \\
Ministry of Science and Technology of the Republic of Slovenia, \\
CICYT, Spain, AEN96--1661 and AEN96-1681,  \\
The Swedish Natural Science Research Council,      \\
Particle Physics and Astronomy Research Council, UK, \\
Department of Energy, USA, DE--FG02--94ER40817. \\

\newpage

\clearpage
\newpage

\setlength{\extrarowheight}{3pt}
\begin{table}[htb]
\begin{center}
\begin{tabular}{|c|c|c|}
  \hline
Selection &\multicolumn{2}{c|}{\stq\; and \sbq: $\dm > 10\:\GeVcc$
      $\sqrt{s}$=183~GeV}\\ 
                             & Data$_{183}$ & MC$_{183}$  \\ \hline \hline
\first\; and \secnd\; step            & 2871 & 2682 $\pm$ 14   \\ \hline
\third\; step  (DLA1) $>$ 0.9         & 98   & 100  $\pm$ 4   \\ 
\third\; step  (DLA2) $>$ 0.          & 27   & 28   $\pm$ 3   \\ \hline
$P_{T_{miss}}\geq $ 12~\GeVc          & 21   & 18   $\pm$ 2   \\ \hline
$E_{jet1} \leq $ 60 GeV               &      &                \\ 
$E_{emjet1}/E_{jet1}\leq$ 0.6         & 7    & 4.4  $\pm$ 0.3   \\ 
$E_{emjet2}/E_{jet2}\leq$ 0.6         &      &          \\ \hline
$20^o \leq \theta_{jets} \leq 160^o$  & 4    & 3.7  $\pm$ 0.3   \\ \hline
$P_{iso} \leq $20~\GeVc               & 3    & 2.9  $\pm$ 0.3   \\ \hline
$|cos \theta_{thrust} | \leq 0.9 $    & 2    & 2.3  $\pm$ 0.3   \\ \hline
$E_{T_{charged}}^{jet 2} \geq $ 2 GeV & 1    & 2.2  $\pm$ 0.3   \\ \hline
Visible mass$\leq$70~\GeVcc           & 1    & 1.8  $\pm$ 0.2   \\ \hline     
$<E_{charged}> \leq$ 4 GeV            & 1    & 1.4  $\pm$ 0.2   \\ \hline
\end{tabular}
\caption{\small Fourth step of the event selection for two-body decays 
  of stop and sbottom in the non-degenerate mass case at $\sqrt{s}$ = 183~GeV. 
  Data$_{183}$ and MC$_{183}$ indicate data and Monte Carlo at $\sqrt{s}$ = 183~GeV.
  DLA1 and DLA2 denote the first and second discriminating linear analysis
  as explained in the text. 
  $P_{T_{miss}}$ stands for the total missing momentum, $E_{jet1}$ ($E_{jet2}$)
  denotes the energy of the (next to) leading jet, $E_{emjet1}$ ($E_{emjet2}$)
  denotes the total electromagnetic energy of the (next to) leading jet,
  $\theta_{jets}$ are the polar angles of the jets, $P_{iso}$ is the momentum
  of the most isolated charged particle, $\theta_{thrust}$ denotes the polar
  angle of the thrust axis and $E_{T_{charged}}^{jet 2} $ is the total tranverse
  energy of the next to leading jet taking into account charged particles
  only.
  $\Delta M$ represents
  the mass difference between the squark and the LSP.
  The case $\dm > 10\:\GeVcc$ corresponds to the non-degenerate mass case.
  The errors on the Monte Carlo are statistical only.}
\label{tab:stsb:sel4-183}
\end{center} 
\end{table}

\setlength{\extrarowheight}{3pt}
\begin{table}[htb]
\begin{center}
\begin{tabular}{|c|c|c|}
  \hline
Selection &\multicolumn{2}{c|}{\stq\; and \sbq: $\dm > 10\:\GeVcc$ $\sqrt{s}$=189 GeV}\\ 
                              & Data$_{189}$ & MC$_{189}$        \\ \hline \hline
\first\; and \secnd\; step    & 6507 & 6659 $\pm$ 12   \\ \hline
\third\; step  (DLA1) $>$ 0.3 &  130 &  125 $\pm$  3   \\ 
\third\; step  (DLA2) $>$ 0.4 &   22 &   24 $\pm$  2   \\ \hline
$R_{30} > $  0.80             &   15 & 12.9 $\pm$  1.1 \\ \hline
$R_{20} > $  0.95             &   12 &   11 $\pm$  0.9 \\ \hline
$P^{leading} < $  25 $\GeVc$  &    7 &  7.6 $\pm$  0.9 \\ \hline
$E_{em2}/E(jet2) \leq $ 0.2   &    5 &    7 $\pm$  0.9 \\ \hline
${\mathcal P}_{comb}\geq$-1   &    2 &  2.2 $\pm$  0.4 \\ 
for \sbq \; only              &      &                 \\ \hline
\end{tabular}
\caption{\small Fourth step of the event selection for two-body decays 
  of stop and sbottom in the non-degenerate mass case at $\sqrt{s}$ = 189~GeV. 
  Data$_{189}$ and MC$_{189}$ indicate data and Monte Carlo at $\sqrt{s}$ = 189~GeV.
  DLA1 and DLA2 denote the first and second discriminating linear analysis
  as explained in the text.
  $R_{30}$ ($R_{20}$)
  denotes the fraction of the total energy out of the cones of $30^o$ and 
  $150^o$ ($20^o$ and $160^o$) centered on the beam axis.
  $P^{leading}$
  denotes the momentum of the leading particle;
  $E_{em2}$
  denotes the total electromagnetic energy of the next to leading jet.
  ${\mathcal P}_{comb}$ is a b-tagging
  probability as explained in the text. $\Delta M$ represents
the mass difference between the squark and the LSP.
The case $\dm > 10\:\GeVcc$ corresponds to the non-degenerate mass case.
The errors on the Monte Carlo are statistical only.}
\label{tab:stsb:sel4}
\end{center} 
\end{table}

\setlength{\extrarowheight}{3pt}
\begin{table}[htb]
\begin{center}
\begin{tabular}{|c||c|c||c|c|}
\hline
Selection & \multicolumn{4}{|c|}{\stq:  $\dm \leq 10\:\GeVcc$  }  \\ 
  & Data$_{183}$ & MC$_{183}$  & Data$_{189}$ & MC$_{189}$   \\   \hline \hline
\first\; and \secnd\; step       &575&528$\pm$7& 1613 & 1567$\pm$45     \\ \hline
\third\; step  (DLA) $>$ 0.3     &44&45$\pm$2&  139 & 134$\pm$10     \\ \hline
oblateness $\geq$0.1             &40&38$\pm$2&  115 & 106$\pm$5     \\ \hline
$R_{30} \geq 0.9 $               &24&25$\pm$1&  76  & 79$\pm$4     \\ \hline
$R_{20}  \geq 0.985 $            &20&22$\pm$1&  65  & 68$\pm$4     \\ \hline
$P_{tt} \leq 30\:\GeVc$          &8&13$\pm$1&  29  & 40$\pm$4     \\ \hline
acoplanarity$_{thrust} \geq 20\dgree$ &1&2.6$\pm$0.5&  8   & 8.1$\pm$1.6     \\ \hline
$\cos({acoplanarity}) \geq -0.85$        &1&0.98$\pm$0.27&  3   & 3.3$\pm$0.8     \\ \hline
\end{tabular}
\caption{\small Fourth step of the event selection for two-body decays of
stop squarks in the degenerate mass case at $\sqrt{s}$ = 183~GeV and
$\sqrt{s}$ = 189~GeV. 
Data$_{183}$ and MC$_{183}$ (Data$_{189}$ and MC$_{189}$)
indicate data and Monte Carlo at $\sqrt{s}$ = 183~GeV ($\sqrt{s}$ = 189~GeV).
$R_{30}$ ($R_{20}$) denotes the fraction of the total energy out of
the cones of $30^o$ and $150^o$ ($20^o$ and $160^o$) centered on the beam axis and 
acoplanarity$_{thrust}$ 
the acoplanarity
angle with respect to the thrust axis.
For the other variables see the text as well as in
Tables~\ref{tab:stsb:sel4-183}
and~\ref{tab:stsb:sel4}.
DLA stands for discriminating linear analysis. 
$\Delta M$ represents
the mass difference between the squark and the LSP.
The case $ \dm \leq 10\:\GeVcc $ corresponds to the degenerate mass case.
The errors on the Monte Carlo are statistical only.}
\label{tab:stop:sel4}
\end{center}
\end{table}

\setlength{\extrarowheight}{3pt}
\begin{table}[htb]
\begin{center}
\begin{tabular}{|c|c|c|}
\hline
Selection &\multicolumn{2}{|c|}{\sbq:   $\dm \leq 10\:\GeVcc$ }  \\ 
          & Data$_{183}$ & MC$_{183}$        \\ \hline \hline
\first\; and \secnd\; step              & 747   & 629 $\pm$ 8   \\ \hline
\third\; step  (DLA) $>$ 0.7            & 70    & 52 $\pm$ 2    \\ \hline
 $E_{tot} \leq $ 40 GeV                 & 42    & 34 $\pm$ 2    \\ \hline
 $E_{T_{charged}}^{jet1} \geq $ 2 GeV   & 32    & 27 $\pm$ 2    \\ \hline    
 $20^o\leq\theta_{jets}\leq 160^o$      & 26    & 25 $\pm$ 2    \\ \hline
 $E_T^{jet1} \geq$ 5 GeV                & 10    & 14 $\pm$ 1    \\ \hline
 acoplanarity$_{thrust} \geq 20^o $     & 1     & 3 $\pm$ 0.6   \\ \hline
 $E_{T_{charged}}^{jet 2} \geq $ 1 GeV  & 1     & 2.5 $\pm$ 0.5 \\ \hline
 $E_{T_{charged}}^{jet2} \geq $ 2 GeV   & 1     & 1.6 $\pm$ 0.4 \\ \hline
 oblateness $\leq$0.36                  & 1     & 1.1 $\pm$ 0.3 \\ \hline
\end{tabular}
\caption{\small Fourth step of the event selection for two-body sbottom 
decays in the degenerate mass case at $\sqrt{s}$ = 183~GeV. 
Data$_{183}$ and MC$_{183}$ indicate data and Monte Carlo at $\sqrt{s}$ = 183~GeV.
The variables are explained in the caption of Tables
~\ref{tab:stsb:sel4-183}, ~\ref{tab:stsb:sel4} 
and~\ref{tab:stop:sel4}.
DLA stands for discriminating linear analysis. $\Delta M$ represents
the mass difference between the squark and the LSP.
The case $ \dm \leq 10\:\GeVcc $ corresponds to the degenerate mass case.
The errors on the Monte Carlo are statistical only.}
\label{tab:sbot:sel4-183}
\end{center}
\end{table}

\setlength{\extrarowheight}{3pt}
\begin{table}[htb]
\begin{center}
\begin{tabular}{|c|c|c|}
\hline
Selection &\multicolumn{2}{|c|}{\sbq:   $\dm \leq 10\:\GeVcc$ }  \\ 
          & Data$_{189}$ & MC$_{189}$        \\ \hline \hline
\first\; and \secnd\; step           & 5307 & 5644$\pm$106     \\ \hline
\third\; step  (DLA) $>$ 0           &   19 &   26$\pm$7      \\ \hline
$R_{20} \geq $ 0.98                  &   14 &   24$\pm$3      \\ \hline
$R_{20} \times $ $P_T^{miss} \geq 1\:\GeVc$  &   12 &   16$\pm$3      \\ \hline    
$P_{tt} \geq 4\:\GeVc$               &    7 &  9.7$\pm$2      \\ \hline
$\cos({acoplanarity}) \geq -0.98$       &    3 &  3.5$\pm$1      \\ \hline
${\mathcal P}_{comb}\geq$-1          &    1 &  2.3$\pm$0.8    \\ \hline
\end{tabular}
\caption{\small Fourth step of the event selection for two-body sbottom 
decays in the degenerate mass case at $\sqrt{s}$ = 189~GeV. 
Data$_{189}$ and MC$_{189}$ indicate data and Monte Carlo at $\sqrt{s}$ = 189~GeV.
The variables are explained in the caption of Tables
~\ref{tab:stsb:sel4-183},
\ref{tab:stsb:sel4} and~\ref{tab:stop:sel4}.
DLA stands for discriminating linear analysis.
$\Delta M$ represents
the mass difference between the squark and the LSP.
The case $ \dm \leq 10\:\GeVcc $ corresponds to the degenerate mass case.
The errors on the Monte Carlo are statistical only.}
\label{tab:sbot:sel4}
\end{center}
\end{table}

\clearpage
\newpage

\begin{table}[htbp]
\begin{center}
\begin{tabular}{|c|c|c|}
\hline
Squark & Data$_{183}$ &  MC$_{183}$ \\
\hline
\stq & 2 & 2.4$\pm$0.3(stat)$^{+0.2}_{-0.2}$(syst) \\
\hline
\sbq & 2 & 2.6$\pm$0.4(stat)$^{+0.3}_{-0.2}$(syst) \\
\hline
\end{tabular}
\caption{\small Number of candidates and expected background in the 
  search for two-body decays of stop and sbottom when performing the 
  $\prime \prime$or$\prime \prime$  
  of the analyses in the de generate and non-degenerate mass case at
   $\sqrt{s}$ = 183~GeV.
 Data$_{183}$ and MC$_{183}$ indicate data and Monte Carlo at $\sqrt{s}$ = 183~GeV.
There are candidates
in common in
the stop and sbottom analyses.}
\label{tab:stsb:bkg-183}
\end{center}
\end{table}

\begin{table}[htb]
\begin{center}
\begin{tabular}{|c|c|c|}
\hline
Squark & Data$_{189}$ &  MC$_{189}$ \\
\hline
\stq & 8 & 9.3$\pm$1.2(stat)$^{+0.9}_{-0.6}$(syst) \\
\hline
\sbq & 3 & 4.4$\pm$0.9(stat)$^{+0.6}_{-0.3}$(syst) \\
\hline
\end{tabular}
\caption{\small Number of candidates and expected background in the 
  search for two-body decays of stop and sbottom when performing the 
  $\prime \prime$or$\prime \prime$ 
  of the analyses in the degenerate and non-degenerate mass case at
  $\sqrt{s}$ = 189~GeV.
 Data$_{189}$ and MC$_{189}$ indicate data and Monte Carlo at $\sqrt{s}$ = 189~GeV.
There are candidates
in common in
the stop and sbottom analyses.}
\label{tab:stsb:bkg}
\end{center}
\end{table}

\clearpage
\newpage

\newpage

\newpage

\newpage

\newpage

\newpage

\newpage


\begin{thebibliography}{99}

\bibitem{DELLIM}
DELPHI Collaboration: P. Abreu \etal, \EJC{6}{99}{385}.
\bibitem{LEPLIM}
ALEPH Collaboration: R. Barate \etal,
\PLB{434}{98}{189}; \\
ALEPH Collaboration: R. Barate \etal,
\PLB{469}{99}{303}; \\
ALEPH Collaboration: R. Barate \etal,
CERN-EP/2000-85; \\
L3 Collaboration, M. Acciari \etal, \PLB{471}{99}{308}; \\            
OPAL Collaboration: K. Ackerstaff \etal, \PLB{396}{97}{301}; \\
OPAL Collaboration: K. Ackerstaff \etal, \ZPC{75}{97}{409}; \\
OPAL Collaboration: G. Abbiendi \etal, \PLB{456}{99}{95}.
\bibitem{MSSM}
P. Fayet and S. Ferrara, \PR{32}{77}{249}; \\
H.P. Nilles, \PR{110}{84}{1}; \\
H.E. Haber and G. L.Kane, \PR{117}{85}{75}.
\bibitem{CHLIM}
ALEPH Collaboration: R. Barate \etal, \EJC{2}{98}{417}; \\
ALEPH Collaboration: R. Barate \etal, \EJC{11}{99}{193}; \\
DELPHI Collaboration: P. Abreu \etal, \EJC{1}{98}{1}; \\
L3 Collaboration: M. Acciarri \etal, Phys. Lett. {\bf B472} (2000) 420; \\
OPAL Collaboration: K. Ackerstaff \etal, \EJC{2}{98}{213};\\
OPAL Collaboration: K. Ackerstaff \etal, E. Phys. J. {\bf C14} (2000) 187.
\bibitem{DELPHIPER1}
DELPHI Collaboration: P. Aarnio \etal, \NIMA{303}{91}{233}.
\bibitem{DELPHIPER2}
DELPHI Collaboration: P. Abreu \etal, \NIMA{378}{96}{57}.
\bibitem{comb}
DELPHI Collaboration: P. Abreu \etal, \ZPC{70}{96}{531};\\
DELPHI Collaboration: P. Abreu \etal, E. Phys. J. {\bf C10} (1999) 415.
\bibitem{JETSET}
T. Sj\"ostrand, \CPC{82}{94}{74}. 
\bibitem{DELPHIDATA}
DELPHI Collaboration: P. Abreu \etal, \ZPC{73}{96}{11}.
\bibitem{SPRING}
S. Kawabata, \CPC{41}{86}{127}.
\bibitem{BEENAKKER}
W.~Beenakker, R.~Hopker, M.~Spira, and P.M.~Zerwas, \PLB{349}{95}{463}.
\bibitem{SUSYGEN}
S.Katsanevas and P.Morawitz, \CPC{112}{98}{227}.
\bibitem{EXCALIBUR}
F.A.~Berends, R.~Pittau, R.~Kleiss, \CPC{85}{95}{437}.
\bibitem{TWOGAM}
 T.Alderweirwld \etal, CERN-OPEN-2000-141
\bibitem{BDK} F.A. Berends, P.H. Daverveldt, R. Kleiss,
Monte Carlo Simulation of Two-Photon
Processes, \CPC{40}{86}{271}.
\bibitem{DURHAM} S. Catani, Yu.L. Dokshitzer, M. Olson, G. Turnock
and B.R. Webber, \PLB{269}{91}{432}.
\bibitem{discri} R.A. Fisher, {\it The use of multiple measurements in 
taxonomic problems}, Annals of Eugenics, {\bf 7} (1936).
\bibitem{stoplep1}
OPAL Collaboration: R. Akers \etal, \PLB{337}{94}{207}.
\bibitem{stopd0}
CDF Collaboration: {\it Search for Scalar Top and Scalar
Bottom Quarks in $p\bar p$ collisions at $\sqrt s = 1.8$ TeV}, hep-ex/9910049,
submitted to Phys.Rev.Lett.
\end{thebibliography}
\end{document}